# A BIG-DATA DRIVEN FRAMEWORK TO ESTIMATING VEHICLE VOLUME BASED ON MOBILE DEVICE LOCATION DATA

Mofeng Yang[1], Weiyu Luo[2], Mohammad Ashoori[3], Jina Mahmoudi[4], Chenfeng Xiong[5*], Jiawei Lu[6], Guangchen Zhao[7], Saeed Saleh Namadi[8], Songhua Hu[9] and Aliakbar Kabiri[10]

1. Ph.D. (mofeng@umd.edu)
2. Graduate Research Assistant (wyl@umd.edu)
3. Graduate Research Assistant (mashoori@umd.edu)
4. Ph.D., P.E., Research Scientist (zhina@umd.edu)
5. Assistant Professor (Chenfeng.Xiong@Villanova.edu), *Corresponding Author
6. Graduate Research Assistant (jiaweil9@asu.edu)
7. Graduate Research Assistant (gczhao@umd.edu)
8. Graduate Research Assistant (saeed@umd.edu)
9. Graduate Research Assistant (hsonghua@umd.edu)
10. Graduate Research Assistant (kabiri@umd.edu)

1-4, 7-10: Maryland Transportation Institute (MTI), Department of Civil and Environmental Engineering, 1173 Glenn Martin Hall, University of Maryland, College Park MD 20742, USA.
5: Department of Civil and Environmental Engineering, College of Engineering, Villanova University, Villanova, PA 19085, USA
6. School of Sustainable Engineering and the Built Environment, Arizona State University, Tempe, AZ 85281, USA

Words Count: 4,623 + 2 Tables (250*2) = 5,123

Submission Date: 07/31/2022

# ABSTRACT

Vehicle volume serves as a critical metric and the fundamental basis for traffic signal control, transportation project prioritization, road maintenance plans and more. Traditional methods of quantifying vehicle volume rely on manual counting, video cameras, and loop detectors at a limited number of locations. These efforts require significant labor and cost for expansions. Researchers and private sector companies have also explored alternative solutions such as probe vehicle data, while still suffering from a low penetration rate. In recent years, along with the technological advancement in mobile sensors and mobile networks, Mobile Device Location Data (MDLD) have been growing dramatically in terms of the spatiotemporal coverage of the population and its mobility. This paper presents a big-data driven framework that can ingest terabytes of MDLD and estimate vehicle volume at a larger geographical area with a larger sample size. The proposed framework first employs a series of cloud-based computational algorithms to extract multimodal trajectories and trip rosters. A scalable map matching and routing algorithm is then applied to snap and route vehicle trajectories to the roadway network. The observed vehicle counts on each roadway segment are weighted and calibrated against ground truth control totals, i.e., Annual Vehicle-Miles of Travel (AVMT), and Annual Average Daily Traffic (AADT). The proposed framework is implemented on the all-street network in the state of Maryland using MDLD for the entire year of 2019. Results indicate that our proposed framework produces reliable vehicle volume estimates and also demonstrate its transferability and the generalization ability.

# 1. INTRODUCTION

Vehicle volume measures the amount of traffic traveling through a roadway segment given a specific period of time. It serves as a critical metric and the fundamental basis for various transportation applications including traffic signal control, transportation project prioritization and road maintenance plan. Traditional methods to quantify vehicle volume rely on manual counting, video cameras, and loop detectors at a limited number of locations, a practice that requires significant human labor and a high cost for expansions (*1-5*). Researchers and private sector companies have also explored alternative solutions such as probe vehicle data, while still suffering from the low penetration rate issue (*6-10*).

In the past two decades, along with the technological advancement in mobile sensors and mobile networks, mobile device location data (MDLD) have been growing dramatically in terms of coverage and size, with broader spatiotemporal coverage of the population and its mobility. A series of research studies have demonstrated the usefulness of MDLD for enhancing the traditional travel survey and have revealed its potential to substitute surveys (*11, 12*). At the same time, obtaining travel statistics solely based on MDLD is also worth investigating to reduce human labor and cost. However, MDLD do not include any ground truth information such as trip origins and destinations, travel modes, and trip purposes, which requires computational algorithms to be developed and validated against the existing travel surveys. More importantly, unlike travel surveys which collect information from representative samples to obtain population-representative statistics, MDLD contain all available mobile devices with uneven data quality.

This study was conducted as part of the *Vulnerable Road User Density Exposure Dashboard project* (*https://mti.umd.edu/sdi*) - an interactive dashboard that utilizes MDLD to provide data and insights on multimodal volume and safety risk exposure of vulnerable road users (e.g., pedestrians, bicycles) at intersections and roadway segments within Maryland. In this study, we present a big-data driven framework that ingests terabytes of MDLD and estimates vehicle volume for all roadway segments. First, a series of cloud-based computational algorithms are applied—including but not limited to—a trip and tour identification algorithm to mine travel behavior information and a travel mode imputation model that impute multimodal trajectories from MDLD. A map matching and routing algorithm is then applied to snap and route vehicle trajectories to the roadway network. The observed vehicle counts on each roadway segment are weighted to match the Annual Vehicle Miles of Travel (AVMT) by county, urban/rural status, and functional classes. Further, a random forest regression model is used to calibrate the weighted vehicle volume against the Annual Average Daily Traffic (AADT) acquired from loop detectors. The proposed framework is implemented on the all-street network in the state of Maryland using MDLD data for the entire year of 2019.

# 2. LITERATURE REVIEW

## 2.1. Application of Mobile Device Location Data in Transportation Research

The appearance of MDLD in the transportation industry started in the 1990s. Since the mid-1990s, researchers began installing Global Positioning System (GPS) data loggers in vehicles to supplement travel surveys (*13-15*). With high-frequency in-vehicle GPS data, this approach can significantly improve the accuracy of travel surveys by recording the exact origin and destination as well as the departure and arrival times. However, only a small number of vehicles can be sampled with this technique, a drawback limiting its capability. Similarly, the wearable GPS, which was introduced in the early 2000s, allowed respondents to report non-vehicle travel modes while still suffering from small sample size issues (*16, 17*). In the past decade, private sector entities such as INRIX and RITIS also started to incorporate the probe vehicle data into their commercial products (*18-21*). Nonetheless, the low penetration rate (i.e., 2%-10%) of the commercial probe vehicle data remains the core challenge with respect to drawing the whole picture of travel patterns.

As mentioned above, despite having high precision, traditional MDLD usually suffer from small-sample-size issues, which significantly limits the usefulness of the data. Since mobile devices, such as smartphones and tablets, have become more popular, MDLD generated from these devices have a greater potential for being used in transportation applications. These new types of MDLD, namely cellular data and Location-Based Service (LBS) data, offer a more extensive spatiotemporal coverage and a larger sample size. The cellular data are generated through communication between cellphones and cell towers (*22*) and can be further categorized into Call Detail Record (CDR) and sightings (*11*). The CDR data can only capture the cell tower location, whereas the sightings provide the exact latitude and longitude values. Both types of cellular data have been widely applied to research topics such as travel behavior, human mobility, and social networks in the past two decades (*23-31*). Despite the large volume of data, cellular data are limited by their spatial and temporal resolution, which is determined by the density of cell towers and users' cellphone usage levels (*32*). On a positive note, however, cellular data require less advanced phones and can raise fewer user privacy concerns. The LBS data provide the exact locations generated when a mobile application updates the device's location with the most accurate sources, based on the existing location sensors such as Wi-Fi, Bluetooth, cellular tower, and GPS (*11, 23-25, 33, 34*). Many applications have been developed using the LBS data. For instance, a recent smartphone-enhanced travel survey conducted in the U.S. used a mobile application, rMove developed by Resource Systems Group (RSG), to collect high-frequency location data and allow the respondents to recall their trips by showing the trajectories in rMove (*35-38*). Additionally, Airsage leveraged LBS data to develop a traffic platform that can estimate traffic flow, speed, congestion and road user sociodemographic for every road and time of day (*39*). Further, the Maryland Transportation Institute (MTI) at the University of Maryland (UMD) developed the COVID-19 Impact Analysis Platform (*https://data.covid.umd.edu*) to provide insight on COVID-19's impact on mobility, health, economy, and society across the U.S. (*40-43*).

**2.2. Vehicle Volume Estimation Methods**

*2.2.1. Estimating Vehicle Volume with Loop Detectors*

Loop detectors are widely used to record traffic volumes and occupancy levels. These sensors are usually buried under the pavements to detect the induction change from the presence of a vehicle. Kwon et al. 2003 developed an algorithm using data from single loop detectors to estimate truck traffic volumes (*1*). The results showed a 5.7% error compared with the ground truth highway data. Loop detector data were also applied together with probe vehicle data to estimate queue length (*44*) and vehicle volume at a city-wide scale (*45*). Although proven to be efficient in estimating vehicle volume, the high installation and maintenance cost of loop detectors limit their capability of being scaled up to cover the entire transportation network. Therefore, loop detector datasets are often incomplete and mostly unavailable at minor arterials and local streets.

*2.2.2. Estimating Vehicle Volume with Probe Vehicle Data*

In the past two decades, MDLD have gained significant attention and have been utilized for estimating various traffic characteristics including vehicle volume. With the development of MDLD, estimating vehicle volume at the city scale became a reality. Probe vehicles can record their trajectory data with high granularity (i.e., 1Hz). Based on the trajectory data obtained from probe vehicles, a wide range of methods can be used by researchers to solve transportation problems. Zhao et al. proposed novel methods to estimate queue length and vehicle volume based on the probability theory without prior information about the penetration rate or queue length distribution (*6*). Guo et al. estimated vehicle volume and queue length at signalized intersections and proposed a new framework to optimize traffic signal control operations (*7*). Sekuła et al. applied several machine learning and neural networks to estimate historical hourly vehicle volume between sparsely located sensors based on the probe vehicle data (*8*). Shockwave theories were also applied to probe vehicle data by a few studies (*9, 10*).

*2.2.3. Estimating Vehicle Volume with Mobile Device Location Data*

Many studies have been conducted focusing on estimating traffic flow and detecting congestion using cellular data (*46, 47*). Xing et al. 2019 utilized CDR with Time Difference of Arrival (TDOA) positioning technique in order to estimate multimodal traffic volumes on different types of urban roadways by identifying three modes of travel – namely, drive alone, carpooling, and bus (*48*). The results showed that compared with the ground truth vehicle volume obtained from License Plate Recognition (LPR) cameras, the mean relative error was in the range of 17.1% to 25.7%, depending on the roadway type. Despite significant advances in positioning techniques, cellular data still suffers from low accuracy issues, whereas LBS data have a noticeable advantage due to utilizing different sources to accurately locate the user – a feature that has resulted in an increased usage of this type of data by researchers and the private sector for estimating vehicle volume. Fan et al. 2019 developed a computing framework alongside a heuristic map matching algorithm to estimate Vehicle Miles of Travel (VMT) and AADT for the state of Maryland using INRIX data. The results showed an $R^2$ of 0.878 when fitting the estimated AADT with the ground truth AADT (*49*). Moreover, a number of state agencies conducted rigorous evaluations of vehicle volume obtained through traditional methods as well as from MDLD obtained by private sector companies. They found the latter to be a promising source for supplementing current surveys and traditional methods (*50*).

## 3. THE BIG-DATA DRIVEN VEHICLE VOLUME ESTIMATION FRAMEWORK

### 3.1. Overview of the Framework

In this study, we propose a big-data driven vehicle volume estimation framework, which offers the capability of efficiently estimating vehicle volume ingested from terabytes of MDLD. Figure 1 shows the proposed framework. The proposed framework is built on Amazon Web Services (AWS). MDLD and all supporting data are stored in Simple Cloud Storage (S3). All algorithms are developed based on Apache Spark, which uses Resilient Distributed Datasets (RDD), and are coded in PySpark using the Elastic MapReduce (EMR) services. In the cloud environment, MDLD are spliced into RDDs given the number of executors (*43, 49*). At the same time, all external data sources (i.e., K-D Tree, network, routing engine) are broadcasted into all executors for master and core nodes. The same algorithms are applied to each RDD along with the broadcasted variables, and the results are aggregated and outputted into S3.

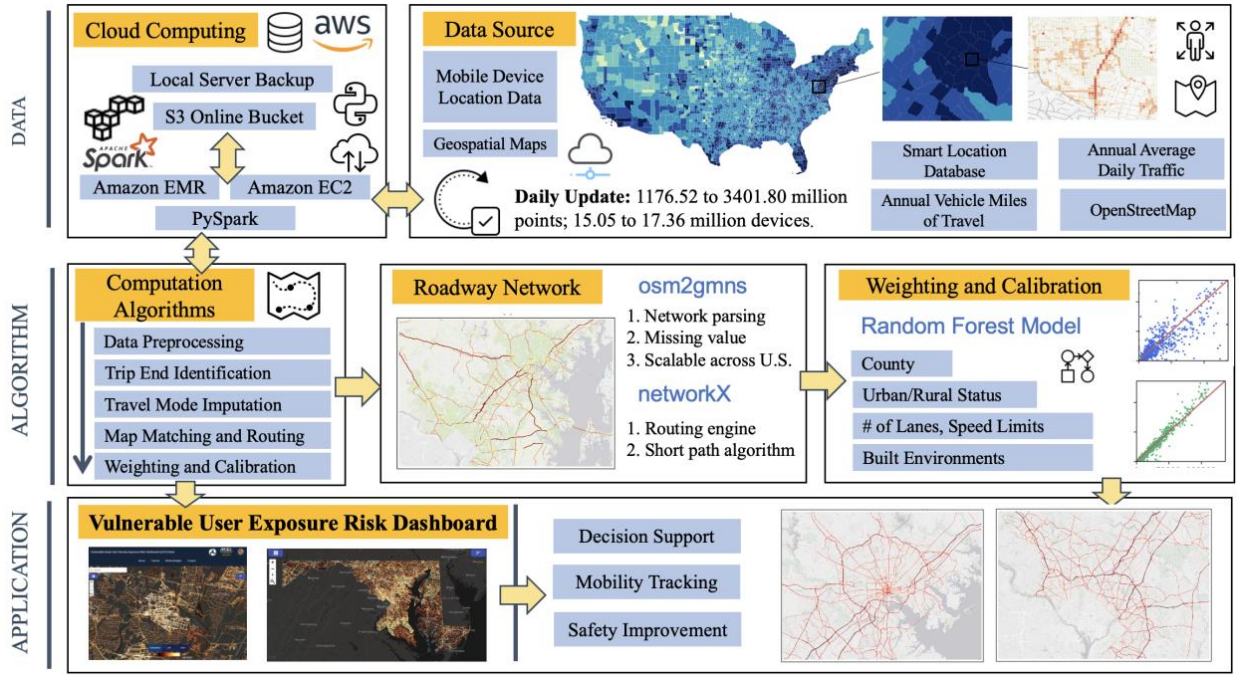

Figure 1. The Big-Data Driven Vehicle Volume Estimation Framework

## 3.2. Trip End Identification and Travel Mode Imputation

Trip is the basic unit of analysis for almost all transportation applications. However, MDLD usually do not contain any trip-related information. Therefore, in this study, a trip end identification algorithm is used to extract trip-level information from the MDLD, including trip start location, trip end location, departure time, and arrival time. Then, a travel mode imputation model is further applied to infer four travel modes–namely, the air, drive, rail, and nonmotorized modes based on heuristic rules and a random forest model. Detailed descriptions of the trip end identification algorithm and the travel mode imputation model can be found in the following references (*12, 51*).

## 3.3. Map Matching and Routing

To ensure flexibility and scalability of our map matching and routing method across the entire U.S., we extract the drivable network from OpenStreetMap (OSM) using the latest open-source Python package *osm2gmns*. The *osm2gmns* package can parse roadway network data from OSM and output networks to csv files in the General Modeling Network Specification (GMNS) format. It provides customized and practical functions to facilitate traffic modeling. Functions include complex intersection consolidation, movement generation, traffic zone creation, short link combination, and network visualization. More details about *osm2gmns* can be found here: https://osm2gmns.readthedocs.io/en/latest/

To match each location sighting to our OSM network, the OSM network is firstly parsed and converted into the routable formats, where roadway segments are represented by links and nodes. With the network topology, we use the *networkX* package to build a shortest path-based routing engine. We then transform the latitude and longitude of the start node and end node for each link to the plane coordinate (in meters), and then calculate link direction (degree) using the arctan value between the two nodes (see Figure 3 for details). The travel direction between consecutive sightings is also calculated. Similar to the method for link direction calculation, the coordinates of each sighting are converted to plane coordinates, then the degree is calculated using the arctan value between consecutive sightings. A spatial index structure, K-Dimensional Tree (K-D Tree), is built using the link geometric nodes (i.e., link nodes). Then, for each sighting, we search all link nodes that are within 100 meters. The 100-meter threshold is selected to balance the algorithm efficacy and the computing speed. If we increase the value, more candidate links will be considered but this will require more computing resources. If we decrease the value, we might not be able to find a candidate link when the observation is sparse. To validate, we calculate the distance between consecutive link nodes using the Maryland OSM network as an example. Results indicate that more than 95% of the link nodes are within 100 meters of their neighbors, as shown in Figure 2. Therefore, using the 100-meter value as the radius for searching candidate nodes is reasonable.

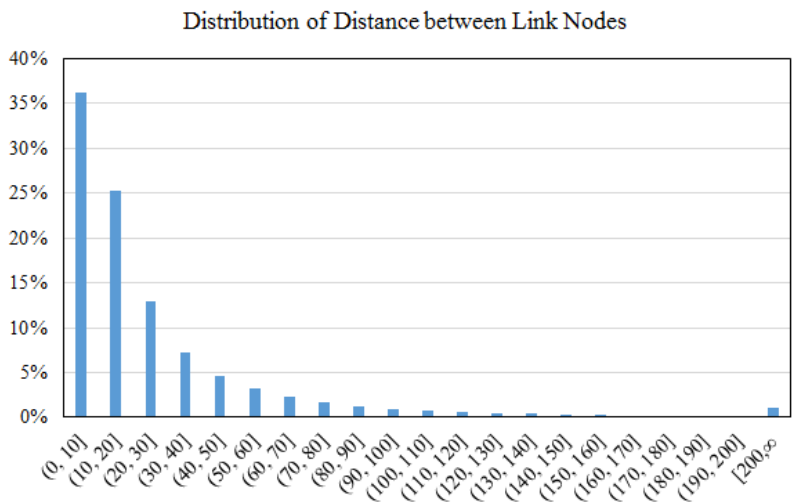

Figure 2. Distribution of Distance between Link Nodes in the OSM Network

As the next step, for each sighting, we compare its travel direction to all candidate links. The closest link with an absolute travel direction difference smaller than 30 degrees will be selected as a valid matched link for the sighting. This 30-degree threshold is selected mainly to avoid the sighting being matched to the link in the opposite direction. In common cases, the degree difference between the travel direction and the link direction should be approximately 0. Here, we use a 30-degree threshold to consider the uncertainty of location accuracy in MDLD. After the matched link for each sighting is found, given the observed link sequence, the routing engine can fill the gap between consecutively observed links and retrieve the complete route. Another layer of reasonable checks is conducted at the routing stage. For each pair of consecutive sightings that are snapped to links, the routed distance is calculated by summing the link length of all the links traveled between the two sightings. Two reasonableness checks are carried out:

(1) If the routed distance is greater than the cumulative distance between the two sightings snapped to links by 2,000 meters or more, we consider the route invalid.
(2) The travel time on these links will be calculated based on the timestamp difference between the two sightings. With the routed distance and travel time, the average travel speed on these links can be calculated. If the speed exceeds 50 m/s (i.e., 112 mph or 180 km/h), we assume that one of the two sightings is matched to the wrong link.

If either of these two violations is observed, we apply a trial-and-error process by removing the latter sighting and performing the routing using the next sighting snapped to the network until it does not violate the 2,000-meter threshold or the 50 m/s threshold (*52*). A simple example of the map matching and routing method is illustrated in Figure 3.

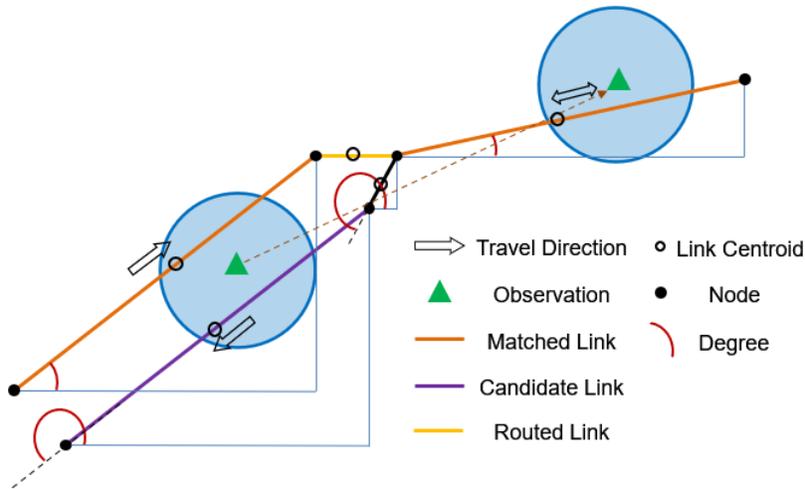

Figure 3. Example of Map Matching and Routing.

### 3.4. Weighting

After map matching and routing, we collect routes for all vehicle trips and aggregate them by links to obtain the observed vehicle volume for each link. Afterward, we develop a link-based weighting method to match the AVMT in the region. We classify each link by county, urban/rural status, and functional classes and calculate the link weight using the formula below:

$$w_{C,u,f} = \frac{AVMT_{C,u,f}}{\sum_{N_C} O_{C,u,f,i}}$$

where $w_{C,u,f}$ represents the weight for links in county $C$, with urban/rural status of $u$, and with functional class $f$; $AVMT_{C,u,f}$ represent the AVMT; and $O_{C,u,f,i}$ represents the observed vehicle volume on link $i$; $N_C$ represents the total number of links in county $C$. For instance, if the study area has 20 counties, 2 urban/rural status and 6 functional classes, then a total of 240 link-based weights will be generated. Subsequently, the weighted vehicle volume for each link can be calculated as:

$$V_{c,u,f,i} = w_{C,u,f} \times O_{c,u,f,i}$$

where $V_{c,u,f,i}$ represents the weighted vehicle volume on link $i$.

### 3.5. Volume Calibration

The weighted vehicle volume is further calibrated to match the ground truth AADT collected from loop detectors at a limited number of locations. In this study, we use the random forest regression to calibrate the weighted vehicle volume against the AADT to obtain the final vehicle volume. During the calibration process, a 10-fold cross-validation (CV) process is used to fine-tune the random forest regression hyperparameters with 90% training data. The fine-tuned models are then applied to the 10% testing data.

## 4. CASE STUDY: THE STATE OF MARYLAND

### 4.1. Data

#### 4.1.1. Mobile Device Location Data and the Study Area

This study used MDLD data obtained from Maryland Transportation Institute (MTI). MTI integrated and cleaned the raw MDLD from multiple data vendors and built a national MDLD data panel that consists of more than 270,000,000 Monthly Active Users (MAU) and represents movements across the nation. (*40-43, 51*). Figure 4 shows the density of location sightings covering locations within and outside of the boundaries of the state of Maryland. In this study, we used all MDLD data that are observed in the state of Maryland for the entire year of 2019. The MDLD is processed on a daily basis and the results are aggregated to produce an annual total result.

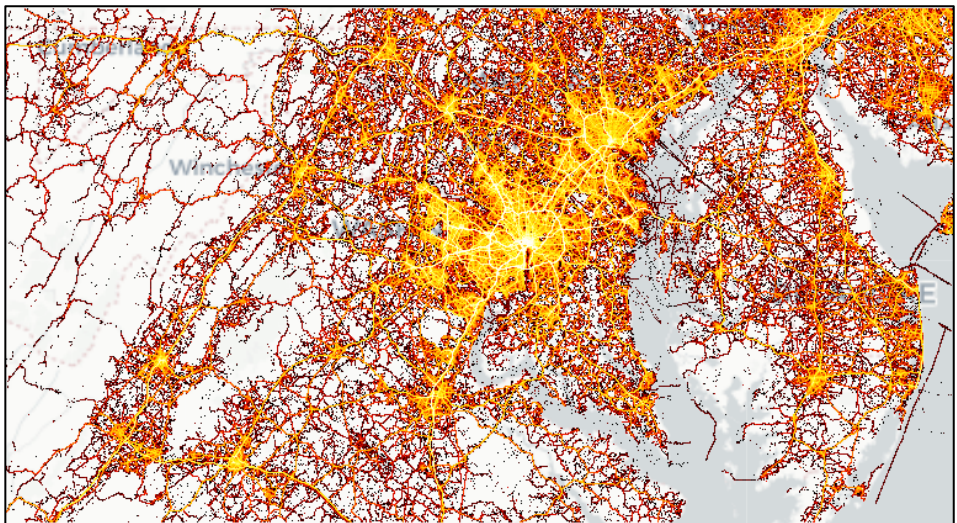

Figure 4. Mobile Device Location Data around the State of Maryland.

*4.1.2. OpenStreetMap Network*

Using the *osm2gmns* package, we extracted a total of 634,516 drivable roadway segments within the state of Maryland. Information about the number of lanes and speed limits was recorded for only 111,835 roadway segments (17.6%) and 84,728 roadway segments (13.4%), respectively. As shown on the left-hand side in Figure 5, the missing values for the number of lanes and speed limits were estimated based on the corresponding values on nearby roadways in the same county, and with the same urban/rural status, and road functional classes. These two variables are further used as features in the vehicle volume calibration model.

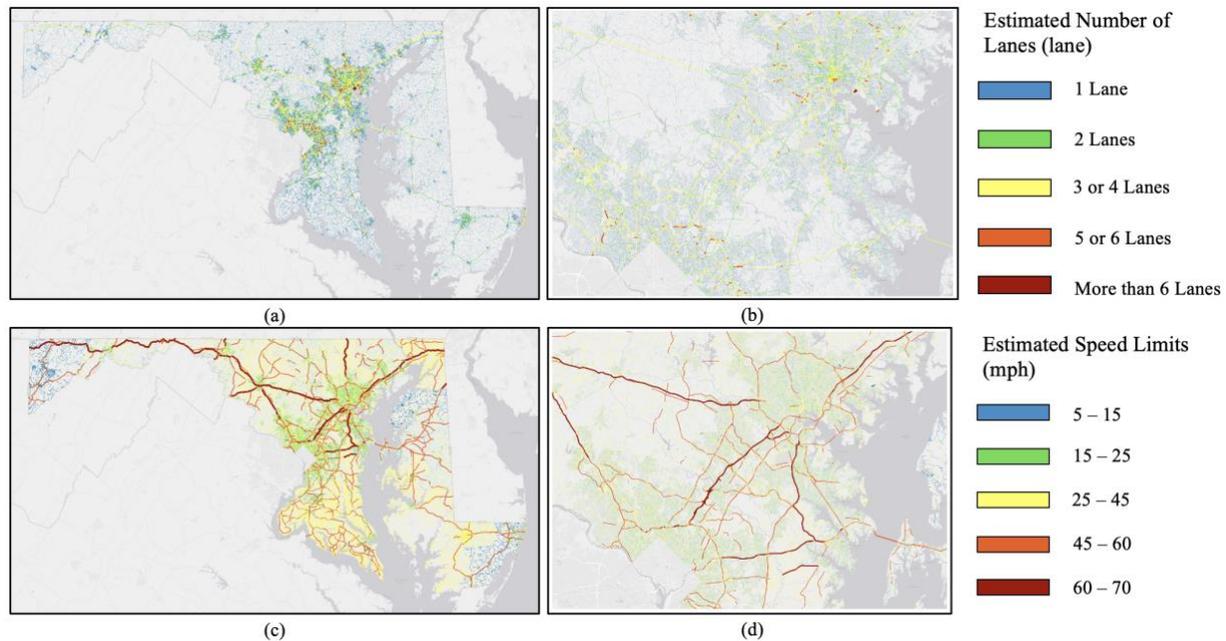

Figure 5. Number of Lanes and Speed Limits in OSM

*4.1.3. Annual Vehicle Miles of Travel Data*

We use the vehicle miles traveled data from the Maryland Department of Transportation State Highway Administration (MDOT SHA) as a control total number to weight observed vehicle volume. Every year, MDOT SHA publishes an annual vehicle miles of travel (AVMT) report by county and functional classification for the state, county, and municipal highway systems. This AVMT report features the current FHWA Functional Classification Codes (1-7) and provides additional classifications (i.e., Urban, Rural, Principal Arterial and Other Freeways and Expressways, and Minor Collector). As discussed in the methodology section, the weights are generated based on county, urban/rural status and functional classes. Here, 23 Maryland counties plus Baltimore City, urban or rural, and two function classes (highway and non-highway) are considered. We map the OSM link type to the FHWA Functional Classification Codes and generated the highway and non-highway classes. More specifically, "motorway", "trunk" and "ramp" are classified as highway (i.e., 1, 2 in FHWA class), and the other types are classified as non-highway (i.e., 3,4,5,6,7 in FHWA class). More details about the AVMT data can be found here: https://www.roads.maryland.gov/mdotsha/Pages/index.aspx?PageId=302

*4.1.4. Annual Average Daily Traffic Data*

We use the AADT also from MDOT SHA to calibrate weighted vehicle volume against the ground truth at a limited number of locations. The AADT data consists of linear and point geometric features which represent the geographic locations and segments of roadway throughout the state of Maryland that include traffic volume metrics such as AADT. More details about the AADT can be found here:https://data.imap.maryland.gov/maps/77010abe7558425997b4fcdab02e2b64/about

*4.1.5. Smart Location Database and Features for Volume Calibration*

The Smart Location Database (SLD) is a nationwide geographic data resource for measuring location efficiency. The SLD is produced by the U.S. Environmental Protection Agency (EPA)'s Smart Growth Program. It provides more than 90 variables on land use and built environment characteristics such as population and employment densities, land use diversity, urban design attributes, destination accessibility, transit accessibility, and socioeconomic/sociodemographic characteristics at the census block group level. Most attributes are available for every census block group in the United States. In this study, we use SLD variables as features in the random forest regression to calibrate weighted vehicle volume to account for the effects of the built environment. The SLD variables used in this study include "*TotEMP*", "*Pct_AO0*", "*D1A*", "*D1C*", "*D3AAO*", "*D3B*", and "*D5AR*":
- TotEMP = total employment;
- Pct_AO0 = percent of zero-car households;
- D1A = gross residential density (housing units per acre) on unprotected land;
- D1C = gross employment density (jobs per acre) on unprotected land;
- D3AAO = network density in terms of facility miles of auto-oriented links per square miles;
- D3B = street intersection density (weighted, auto-oriented intersections eliminated);
- D5AR = jobs within 45 minutes auto travel time, time decay (network travel time) weighted

We also include urban/rural status, county code, link type, number of lanes, and speed limits as features in the calibration process.

**4.2. Results**

*4.2.1. Overall Comparison*

Figure 6 shows the weighting and calibration results for both training and testing sets. The blue dots represent weighted volume comparisons and the green dots represent calibrated vehicle volume comparisons with MDOT SHA AADT. Figure 6 (a) and (b) compares the weighted vehicle volume and calibrated vehicle volume with the MDOT SHA AADT in the training set respectively; Figure 6 (c) and (d) compares the weighted vehicle volume and calibrated vehicle volume with the MDOT SHA AADT in the testing set respectively. As it can be seen from Figure 6 (a), for the training set, the Pearson correlation value and the Root Mean Square Error (RMSE) between the weighted vehicle volume and the ground truth AADT are 0.746 and 7,912, respectively. These values are improved to 0.966 and 2,996 after calibration, as shown in Figure 6 (b). Similarly, for the testing set, the Pearson correlation and RMSE are improved from 0.764 and 7,548, to 0.854 and 5,701 respectively after calibration.

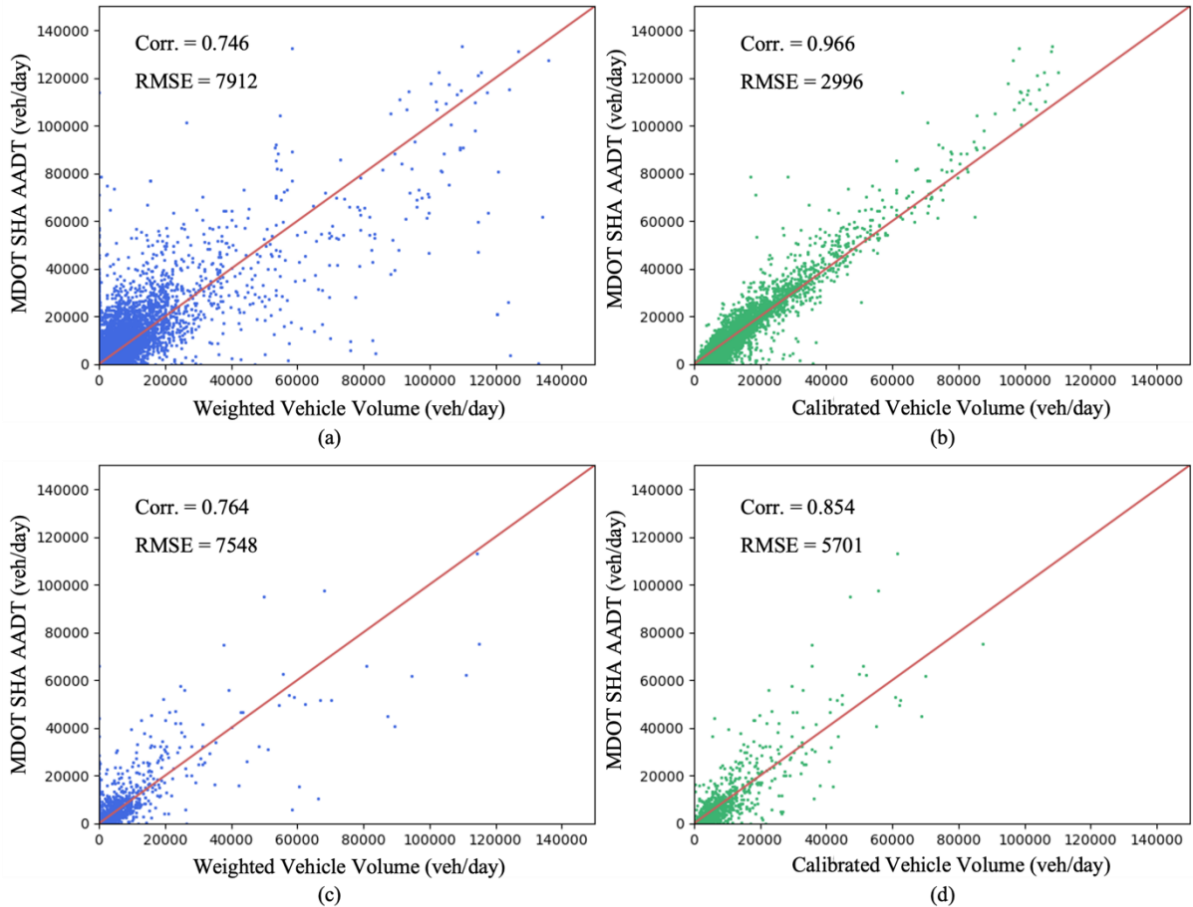

Figure 6. (a) Weighted Vehicle Volume in Training Set; (b) Calibrated Vehicle Volume in Training Set; (c) Weighted Vehicle Volume in Testing Set; (d) Calibrated Vehicle Volume in Testing Set.

*4.2.2. Vehicle Volume Validation by Link Types and Urban/Rural Status*

Figure 7 and Table 1 show the calibrated vehicle volume by link types for both the training and testing sets. For all link types, a good correlation (i.e., over 0.80) can be observed between the calibrated vehicle volume and the ground truth AADT, except for Local Roads and Highway Ramps in the testing set. The results indicate that our proposed framework can accurately estimate vehicle volume on higher-level roadways (i.e., Interstate Highways and Highways, Primary Roads, Secondary Roads), while concurrently maintaining high correlations for lower-level roadways (i.e., Tertiary Roads, Local Roads, Highway Ramps). The relatively weaker performance for the case of lower-level roadways can be attributed to limitations in technology. The MDLD only capture part of the daily trips of a device within the area with mobile network connections and higher-level roadways usually have a better coverage compared to lower-level ones. This variability might also result in capturing more travelers on highways and major arterials. In addition, the LBS data sample is more likely to include the active travelers that make more trips and/or longer-duration trips, such as long-distance travel for leisure or business purposes or long-distance commute which usually happen on interstate highways.

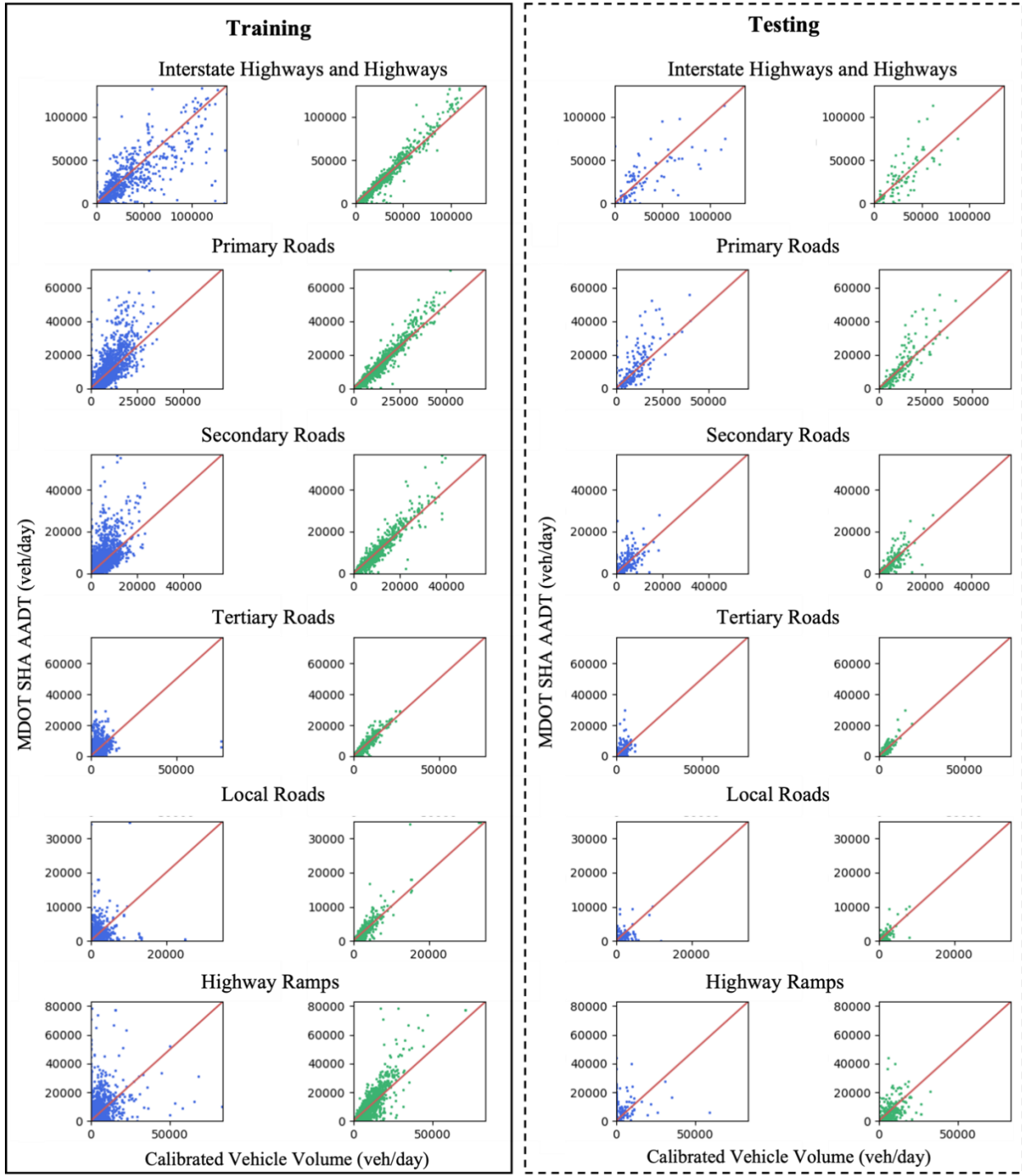

Figure 7. Volume Calibration Results Comparison by Link Type.

Figure 8 and Table 2 show the calibration of vehicle volume by urban/rural status for both the training and testing sets. In summary, for both urban and rural roads, a good correlation (i.e., over 0.80) can be observed between the calibrated vehicle volume and the ground truth AADT, whereas a higher correlation can be observed for urban roads. The relatively weaker performance in rural roadways can also be attributed to the technology limitation mentioned above.

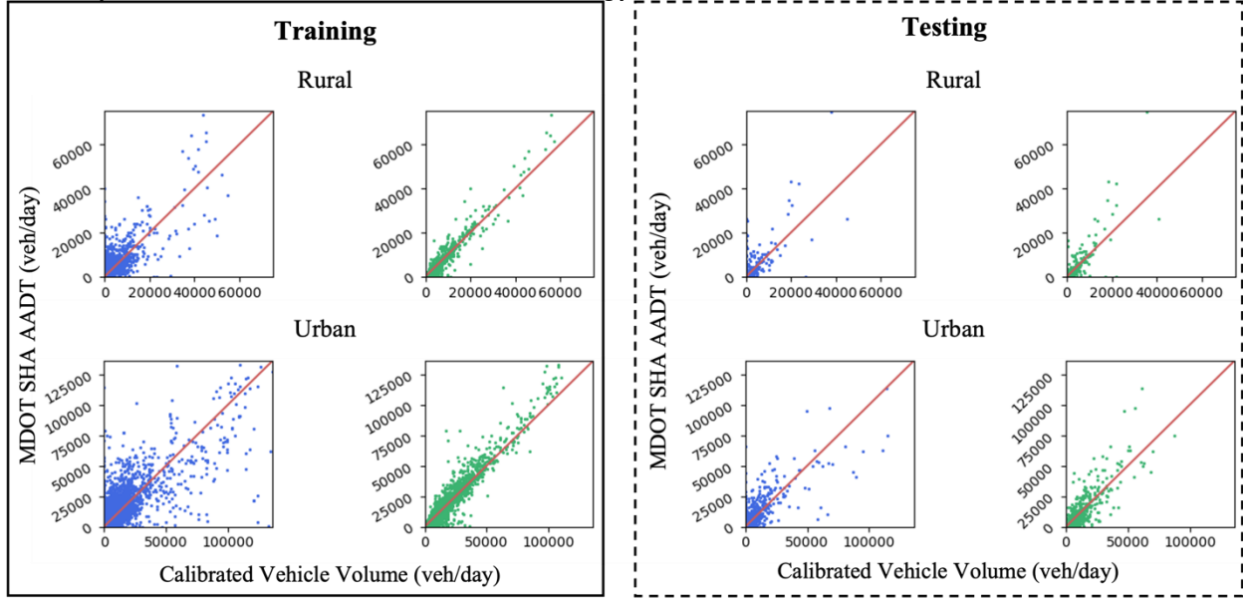

Figure 8. Volume Calibration Results Comparison by Urban/Rural Status.

Table 1. Volume Calibration Results Comparison by Link Type

| Link Type | Training Set | | | | Testing Set | | | |
| --- | --- | --- | --- | --- | --- | --- | --- | --- |
| | Corr. | | RMSE | | Corr. | | RMSE | |
| | Before | After | Before | After | Before | After | Before | After |
| All | 0.746 | 0.966 | 7912 | 2996 | 0.764 | 0.854 | 7548 | 5701 |
| Interstate Highways and Highways | 0.752 | 0.975 | 20081 | 6559 | 0.712 | 0.775 | 19633 | 15246 |
| Primary Roads | 0.699 | 0.971 | 7909 | 2695 | 0.721 | 0.846 | 8665 | 6509 |
| Secondary Roads | 0.627 | 0.960 | 4899 | 1776 | 0.617 | 0.813 | 3667 | 2667 |
| Tertiary Roads | 0.414 | 0.959 | 3486 | 994 | 0.511 | 0.869 | 3090 | 1877 |
| Local Roads | 0.374 | 0.944 | 2474 | 853 | 0.426 | 0.742 | 1701 | 1083 |
| Highway Ramps | 0.242 | 0.866 | 10426 | 4722 | 0.182 | 0.402 | 9119 | 6846 |

Table 2. Volume Calibration Results by Urban/Rural Status.

| Link Type | Training Set | | | | Testing Set | | | |
| --- | --- | --- | --- | --- | --- | --- | --- | --- |
| | Corr. | | RMSE | | Corr. | | RMSE | |
| | Before | After | Before | After | Before | After | Before | After |
| All | 0.746 | 0.966 | 7912 | 2996 | 0.764 | 0.854 | 7548 | 5701 |
| Rural | 0.769 | 0.967 | 3583 | 1442 | 0.727 | 0.826 | 4810 | 4075 |
| Urban | 0.738 | 0.964 | 8913 | 3363 | 0.764 | 0.853 | 8311 | 6179 |

Figure 9 visualizes the calibrated vehicle volume averaged from the entire year of 2019 (represented as AADT) on the all-street network in the state of Maryland. It can be seen that the interstate highway and the highway skeletons can be clearly identified from the map. Major arterials also stand out from the map. Figure 9 (b) zooms into the Washington D.C. area, where I-495, I-270, I-95 and the Baltimore/Washington Parkway are clearly seen. Figure 9(c) zooms into the Baltimore area, where I-395, I-695, I-795, I-95, and I-70 are all captured. Figure 9(d) zooms into Hagerstown, MD, which is a city in Washington County, MD near the border of Pennsylvania. The I-70, I-81, and MD-40 are all captured, demonstrating the ability of our proposed framework to produce reliable results in rural areas.

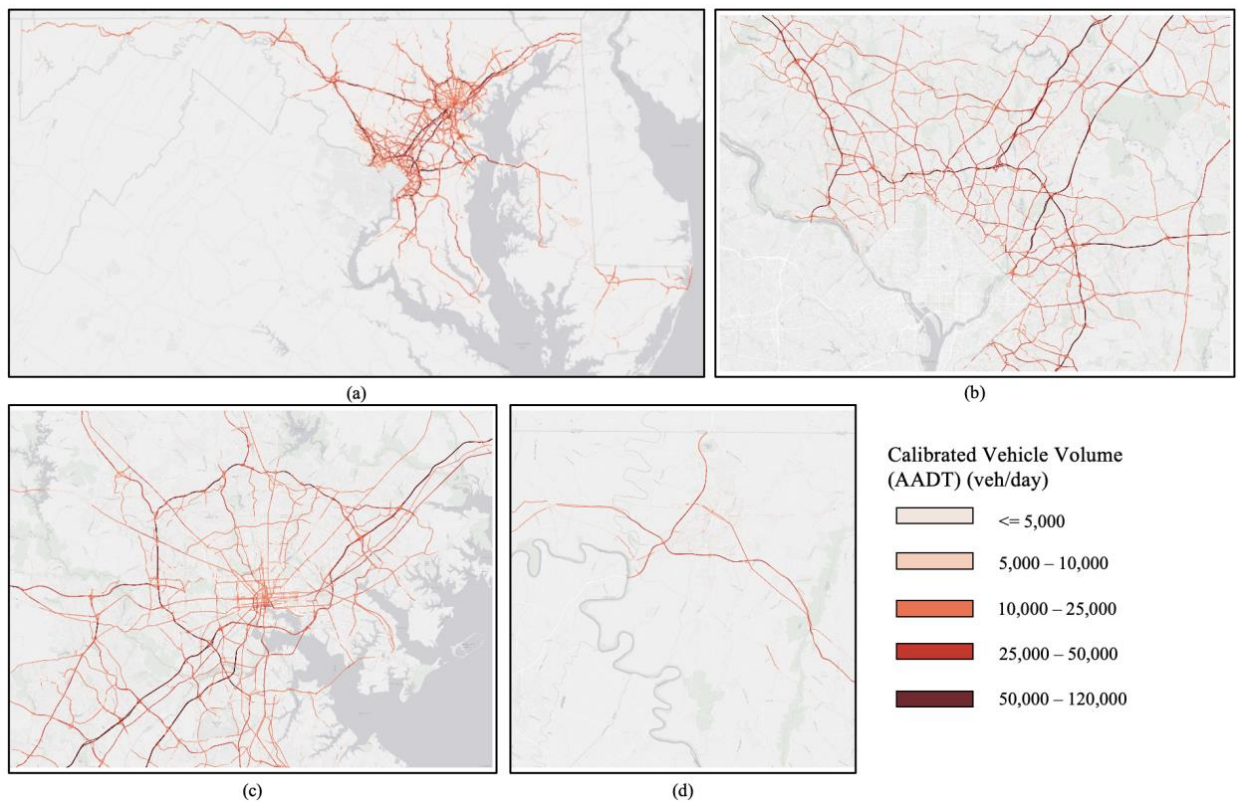

Figure 9. Visualization of Calibrated Vehicle Volume. (a) the State of Maryland; (b) Washington D.C.; (c) Baltimore City; (d) Hagerstown, MD.

## 5. CONCLUSIONS AND DISCUSSIONS

This paper presents a big-data driven framework that is able to ingest terabytes of MDLD and estimate vehicle volume based on MDLD. The proposed framework first employs a series of cloud-based computational algorithms to extract vehicle trajectories. A map-matching and routing algorithm is then applied to snap and route vehicle trajectories to the road network. The observed vehicle counts on each road segment are weighted and calibrated against the control total, i.e., annual vehicle miles traveled (VMT), and data collected from real-world loop detectors. The proposed framework is implemented and validated on the all-street network in the state of Maryland using MDLD data from 2019. After weighting and calibration processes, high correlation and low RMSE values are observed between our vehicle volume estimates and the ground truth data.

The framework proposed in this study and the study findings have practical implications. For instance, estimated vehicle volume based on MDLD can be leveraged in safety risk exposure analysis. In particular, the proposed estimation method can particularly be beneficial for safety risk exposure and crash analysis with respect to vulnerable road users (e.g., pedestrians and bicyclists). Pedestrian and bicyclist exposure data have traditionally been collected through surveys or count collections at sample locations (*53, 54*). In addition to being costly and labor-intensive, these conventional data collection methods are susceptible to subjectivity and may yield inaccurate data. Consequently, high-quality and readily-available pedestrian and bicyclist exposure data are considered as a limitation in safety analysis (*55*). As exposure data are crucial for contextualization of crash analysis and prioritization of safety countermeasures (*53*), utilization of high-quality and consistent exposure data is imperative. When it comes to safety analysis, using MDLD for volume estimation—as performed in this study—provides a tremendous advantage over using data obtained from traditional volume estimation methods. This is due to the potential of the MDLD to produce more reliable exposure data. Employment of such high-fidelity exposure data (i.e., MDLD-estimated volumes) as input for safety and crash analyses can lead to more accurate results and guide data-driven, evidence-based policy decision-making to improve the safety of all road users including the most vulnerable ones.

## ACKNOWLEDGEMENTS
This study was conducted as part of a collaboration among the Maryland Department of Transportation State Highway Administration (MDOT SHA), Maryland Transportation Institute (MTI) at the University of Maryland College Park, and Shock, Trauma and Anesthesiology Research (STAR) Center at the University of Maryland Baltimore through the sponsorship from the Safety Data Initiative from the U.S. Department of Transportation (USDOT).

## CONFLICT OF INTEREST
The authors declare that they have no conflict of interest.

## AUTHOR CONTRIBUTION STATEMENT
The authors confirm contribution to the paper as follows: study conception and design: M.Y., W.L., C.X.; data collection: M.Y., M.A., J.M., G.C., S.S.N.; analysis and interpretation of results: M.Y., J.M., W.L.; methodology support (*osm2gmns*): J.L.; draft manuscript preparation: M.Y., W.L., M.A., J.M., G.C., S.S.N., A.K..

## REFERENCE

1. Kwon, J., Varaiya, P. and Skabardonis, A.. Estimation of truck traffic volume from single loop detectors with lane-to-lane speed correlation. *Transportation Research Record*, *1856*(1), pp.106-117 (2003).
2. Muñoz, L., Sun, X., Horowitz, R., and Alvarez, L.. Traffic density estimation with the cell transmission model. In American Control Conference. Proceedings of the 2003, Vol. 5. IEEE, 3750–3755 (2003).

3. Okutani, I. and Stephanedes, Y. J.. Dynamic prediction of traffic volume through Kalman filtering theory. Transportation Research Part B: Methodological 18, 1 (1984), 1–11
4. Wilkie, D., Sewall, J., and Lin, M.. Flow reconstruction for data driven traffic animation. ACM Transactions on Graphics (TOG) 32, 4 (2013), 89
5. Jia, Z., C. Chen, B. Coifman, and P. Varaiya. The PeMS Algorithms for Accurate, Real-Time Estimates of g-Factors and Speeds from Single Loop Detectors. Proc., IEEE Intelligent Transportation Systems Council Annual Meeting, IEEE, Piscataway, N.J., pp. 536–541 (2001).
6. Zhao, Y., Zheng, J., Wong, W., Wang, X., Meng, Y., & Liu, H. X.. Various methods for queue length and traffic volume estimation using probe vehicle trajectories. Transportation Research Part C: Emerging Technologies, 107, 70–91 (2019). https://doi.org/10.1016/j.trc.2019.07.008
7. Guo, Q., Li, L., & (Jeff) Ban, X.. Urban traffic signal control with connected and automated vehicles: A survey. In Transportation Research Part C: Emerging Technologies (Vol. 101, pp. 313–334) (2019). Elsevier Ltd. https://doi.org/10.1016/j.trc.2019.01.026
8. Sekuła, P., Marković, N., vander Laan, Z., & Sadabadi, K. F.. Estimating historical hourly traffic volumes via machine learning and vehicle probe data: A Maryland case study. Transportation Research Part C: Emerging Technologies, 97, 147–158 (2018). https://doi.org/10.1016/j.trc.2018.10.012
9. Anuar, K., & Cetin, M.. Estimating Freeway Traffic Volume Using Shockwaves and Probe Vehicle Trajectory Data. Transportation Research Procedia, 22, 183–192 (2017). https://doi.org/10.1016/j.trpro.2017.03.025
10. Li, F., Tang, K., Yao, J., & Li, K.. Real-Time Queue Length Estimation for Signalized Intersections Using Vehicle Trajectory Data. Transportation Research Record, 2623(1), 49–59 (2017). https://doi.org/10.3141/2623-06
11. Chen, C., Ma, J., Susilo, Y., Liu, Y., & Wang, M.. The promises of big data and small data for travel behavior (aka human mobility) analysis. *Transportation Research Part C: Emerging Technologies*. 68, 285-299, (2016).
12. Yang, M., Pan, Y., Darzi, A., Ghader, S., Xiong, C. and Zhang, L.. A data-driven travel mode share estimation framework based on mobile device location data. *Transportation*, pp.1-45 (2021).
13. Battelle. Global Positioning Systems for Personal Travel Surveys: Lexington Area Travel Data Collection Test. Final Report. FHWA, U.S. Department of Transportation, (1997).
14. 2000–2001 California Statewide Household Travel Survey. Final Report. NuStats, Austin, Tex (2002).
15. Kansas City Regional Travel Survey. Final Report. NuStats, Austin, Tex, (2004).
16. Mid-Region Council of Governments 2013 Household Travel Survey. Final Report. Westat, Rockville, Md, (2014).
17. 2014 Southern Nevada Household Travel Survey. Final Report. Westat, Rockville, Md, (2015).
18. INRIX Traffic. http://www.inrix.com/, (2020).
19. Haghani, Ali, Masoud Hamedi, and Kaveh Farokhi Sadabadi. I-95 Corridor coalition vehicle probe project: Validation of INRIX data. I-95 Corridor Coalition 9, (2009).
20. Schrank, D., Eisele, B., & Lomax, T.. 2014 Urban mobility report: powered by Inrix Traffic Data (No. SWUTC/15/161302-1), (2015).
21. Cui, Z., Ke, R., Pu, Z., & Wang, Y.. Deep bidirectional and unidirectional LSTM recurrent neural network for network-wide traffic speed prediction. *arXiv preprint* arXiv:1801.02143, (2018).

22. Horak, Ray. *Telecommunications and data communications handbook*. John Wiley & Sons, (2007).
23. Gonzalez, M. C., Hidalgo, C. A., & Barabasi, A. L.. Understanding individual human mobility patterns. *Nature*, 453(7196), 779-782, (2008).
24. Kang, C., Liu, Y., Ma, X., & Wu, L.. Towards estimating urban population distributions from mobile call data. *Journal of Urban Technology*, 19(4), 3-21, (2012).
25. Kang, C., Ma, X., Tong, D., & Liu, Y.. Intra-urban human mobility patterns: An urban morphology perspective. *Physica A: Statistical Mechanics and its Applications*, 391(4), 1702-1717, (2012).
26. Pappalardo, L., F. Simini, S. Rinzivillo, D. Pedreschi, F. Giannotti and A.-L. Barabási. Returners and Explorers Dichotomy in Human Mobility. *Nature communications*. Vol. 6, pp. 8166. (2015).
27. Song, C., T. Koren, P. Wang and A.-L. Barabási. Modelling the Scaling Properties of Human Mobility. *Nature Physics*. Vol. 6, No. 10, pp. 818. (2010).
28. Song, C., Z. Qu, N. Blumm and A.-L. Barabási. Limits of Predictability in Human Mobility. *Science*. Vol. 327, No. 5968, pp. 1018-102. (2010).
29. Çolak, S., A. Lima and M. C. González. Understanding Congested Travel in Urban Areas. *Nature communications*. Vol. 7, pp. 10793. (2016).
30. Bachir, D., Khodabandelou, G., Gauthier, V., El Yacoubi, M. and Puchinger, J. Inferring dynamic origin-destination flows by transport mode using mobile phone data. *Transportation Research Part C: Emerging Technologies*, 101, pp.254-275. (2019).
31. Fekih, M., Bellemans, T., Smoreda, Z., Bonnel, P., Furno, A. and Galland, S.. A data-driven approach for origin–destination matrix construction from cellular network signalling data: a case study of Lyon region (France). *Transportation*, pp.1-32. (2020).
32. Landmark, A.D., Arnesen, P., Södersten, C.J. and Hjelkrem, O.A.. Mobile phone data in transportation research: methods for benchmarking against other data sources. *Transportation*, pp.1-23. (2021).
33. Wang, F., & Chen, C.. On data processing required to derive mobility patterns from passively-generated mobile phone data. *Transportation Research Part C: Emerging Technologies*. 87, 58-74, (2018).
34. Wang, F., Wang, J., Cao, J., Chen, C., & Ban, X. J.. Extracting trips from multi-sourced data for mobility pattern analysis: An app-based data example. *Transportation Research Part C: Emerging Technologies*. 105, 183-202, (2019).
35. Puget Sound Regional Travel Study. Report: Spring 2014 Household Travel Survey. RSG, (2014).
36. In-The-Moment Travel Study. Revised Report. RSG, (2015).
37. Puget Sound Regional Travel Study. Report: 2015 Household Travel Survey. RSG, (2015).
38. 2017 Puget Sound Regional Travel Study. Draft Final Report. RSG, (2017).
39. Airsage. https://www.airsage.com/, (2020).
40. Zhang, L., Darzi, A., Ghader, S., Pack, M.L., Xiong, C., Yang, M., Sun, Q., Kabiri, A. and Hu, S.. Interactive covid-19 mobility impact and social distancing analysis platform. *Transportation Research Record*, p.03611981211043813 (2020).
41. Xiong, C., Hu, S., Yang, M., Luo, W., and Zhang, L.. Mobile device data reveal the dynamics in a positive relationship between human mobility and COVID-19 infections. *Proceedings of the National Academy of Sciences*, 117(44), 27087-27089 (2020a).

42. Xiong, C., Hu, S., Yang, M., Younes, H., Luo, W., Ghader, S. and Zhang, L.. Mobile device location data reveal human mobility response to state-level stay-at-home orders during the COVID-19 pandemic in the USA. *Journal of the Royal Society Interface*, 17(173), p.20200344. (2020b).
43. Hu, S., Xiong, C., Yang, M., Younes, H., Luo, W. and Zhang, L., 2021. A big-data driven approach to analyzing and modeling human mobility trend under non-pharmaceutical interventions during COVID-19 pandemic. *Transportation Research Part C: Emerging Technologies*, *124*, p.102955.
44. Li, J.Q., Zhou, K., Shladover, S.E. and Skabardonis, A.. Estimating queue length under connected vehicle technology: Using probe vehicle, loop detector, and fused data. *Transportation research record*, *2356*(1), pp.17-22 (2013).
45. Meng, C., Yi, X., Su, L., Gao, J. and Zheng, Y.. City-wide traffic volume inference with loop detector data and taxi trajectories. In *Proceedings of the 25th ACM SIGSPATIAL International Conference on Advances in Geographic Information Systems* (pp. 1-10) (2017).
46. Caceres, N., Romero, L. M., Benitez, F. G. and del Castillo, J. M.. Traffic Flow Estimation Models Using Cellular Phone Data. *IEEE Transactions on Intelligent Transportation Systems*, vol. 13, no. 3, pp. 1430-1441 (2012). doi: 10.1109/TITS.2012.2189006
47. Janecek, A., Valerio, D., Hummel, K. A., Ricciato, F. and Hlavacs, H.. The Cellular Network as a Sensor: From Mobile Phone Data to Real-Time Road Traffic Monitoring. *IEEE Transactions on Intelligent Transportation Systems*, vol. 16, no. 5, pp. 2551-2572, doi: 10.1109/TITS.2015.2413215 (2015).
48. Xing, J., Liu, Z., Wu, C., & Chen, S.. Traffic volume estimation in multimodal urban networks using cell phone location data. *IEEE Intelligent Transportation Systems Magazine*, 11(3), 93-104 (2019).
49. Fan, J., Fu, C., Stewart, K., and Zhang, L.. Using big GPS trajectory data analytics for vehicle miles traveled estimation. *Transportation research part C: emerging technologies* 103 (2019): 298-307.
50. Codjoe, Julius, Grace Ashley, and William Saunders. Evaluating cell phone data for AADT estimation. No. FHWA/LA. 18/591, LTRC Project Number: 16-3SA, State Project Number: DOTLT1000110. Louisiana Transportation Research Center, (2018).
51. Zhang, L., Ghader, S., Darzi, A., Pan, Y., Yang, M., Sun, Q., Kabiri, A. and Zhao, G.. Data analytics and modeling methods for tracking and predicting origin-destination travel trends based on mobile device data. *Federal Highway Administration Exploratory Advanced Research Program* (2020).
52. Newson, P. and Krumm, J.. Hidden Markov map matching through noise and sparseness. In *Proceedings of the 17th ACM SIGSPATIAL international conference on advances in geographic information systems* (pp. 336-343) (2009).
53. Sanders, R.L., Frackelton, A., Gardner, S., Schneider, R., Hintze, M. Ballpark Method for Estimating Pedestrian and Bicyclist Exposure in Seattle, Washington: Potential Option for Resource-constrained Cities in an Age of Big Data. *Transportation Research Record*, 2605(1), pp.32–44 (2017).
54. Lee, K., and Sener, I.N. *Emerging Data Mining for Pedestrian and Bicyclist Monitoring: A Literature Review Report*. Safety through Disruption, National University Transportation Center (UTC) Program. (2017). https://safed.vtti.vt.edu/wp-content/uploads/2020/07/UTC-Safe-D_Emerging-Data-Mining-for-PedBike_TTI-Report_26Sep17_final.pdf. Accessed July 10, 2021.

55. *Synthesis of Methods of Estimating Pedestrian and Bicyclist Exposure to Risk at Areawide Levels and on Specific Transportation Facilities*. FHWA-SA-17-041. FHWA, U.S. Department of Transportation, (2017).